\documentclass[aps,twocolumn,floatfix]{revtex4} 

\usepackage{amsmath,amsfonts,amssymb,graphicx,color,times,bbm}

\begin{document}

\bibliographystyle{apsrev}
\newtheorem{theorem}{Theorem}
\newtheorem{corollary}{Corollary}
\newtheorem{definition}{Definition}
\newtheorem{proposition}{Proposition}
\newtheorem{lemma}{Lemma}
\newcommand{\proofend}{\hfill\fbox\\\medskip }
\newcommand{\proof}[1]{{\bf Proof.} #1 $\proofend$}
\newcommand{\nn}{{\mathbbm{N}}}
\newcommand{\rr}{{\mathbbm{R}}}
\newcommand{\cc}{{\mathbbm{C}}}
\newcommand{\zz}{{\mathbbm{Z}}}
\newcommand{\mbp}{\ensuremath{\spadesuit}}
\newcommand{\je}{\ensuremath{\heartsuit}}
\newcommand{\jd}{\ensuremath{\clubsuit}}
\newcommand{\id}{{\mathbbm{1}}}
\renewcommand{\vec}[1]{\boldsymbol{#1}}
\newcommand{\me}{\mathrm{e}}
\newcommand{\mi}{\mathrm{i}}
\newcommand{\md}{\mathrm{d}}
\newcommand{\sg}{\text{sgn}}

\delimitershortfall=-2pt

\title{Measuring entanglement in condensed matter systems}

\author{M.~Cramer$^1$, M.B.~Plenio$^{1,2}$, and H.~Wunderlich$^1$}

\affiliation{$^1$Institut f\"ur Theoretische Physik, Albert-Einstein
Allee 11, Universit\"at Ulm, Germany\\
$^2$Quantum Optics and Laser Science group, Imperial College London, London SW7 2BW, UK}

\begin{abstract}
We show how entanglement may be quantified in spin and cold atom many-body 
systems using standard experimental techniques only. The scheme requires 
no assumptions on the state in the laboratory and a lower bound to the 
entanglement can be read off directly from the scattering cross section of 
Neutrons deflected from solid state samples or the time-of-flight distribution 
of cold atoms in optical lattices, respectively. This removes a major obstacle
which so far has prevented the direct and quantitative experimental study of
genuine quantum correlations in many-body systems: The need for a 
full characterization of
the state to quantify the entanglement contained in it. Instead, the scheme
presented here relies solely on global measurements that are routinely 
performed and is versatile enough to accommodate systems and measurements
different from the ones we exemplify in this work.

\end{abstract}

\maketitle

\date{\today}

\section{Introduction} Interacting quantum many-body systems generally
exhibit correlations between its constituents. At sufficiently low
temperatures, near the ground-state, these correlations possess quantum
mechanical features, namely entanglement. Compared to its classical
counterpart, entanglement is extremely complex. Its full characterization
generally requires the measurement of a number of observables that
grows exponentially with the number of constituents of the quantum
many-body systems. On the one hand, the ability to create entanglement
merely by cooling an interacting quantum many-body system provides the
attractive opportunity of using this entanglement to carry out quantum
information processing tasks such as measurement based quantum computation \cite{mmb_qc} or adiabatic quantum computation \cite{adiabatic_qc}
which gain their power exactly because of the complex structure of entanglement.
However, the very same setting offers significant challenges, as it
is much harder to analyze theoretically and, crucially, experimentally:
While for the few-particle systems that can now be prepared in highly
controlled environments such as ion traps, it is possible to fully
characterize the state in the laboratory by quantum state tomography
\cite{tomography}, the situation in condensed matter systems is far more
challenging. Firstly, the number of subsystems tends to be much larger,
the level of control over states and Hamiltonians is more restricted and
crucially the available measurements for condensed matter systems are much
less general: Local measurements addressing individual constituents are
usually not available and one has to rely on global measurements such as
those obtained in scattering experiments to draw conclusions about the
system. These are of course by no means sufficient to fully characterize
the state in the laboratory. How might one still be able to say something
about the entanglement that is available? One approach could be to, e.g.,
model the system with a certain quantum Hamiltonian and compare it to a
classical model. If the predictions from the quantum model matches the
measurement results while the classical does not, and the simulated quantum
state displays entanglement, one might conclude that the state in the
laboratory is indeed entangled. This however is a fallacy. Consider the
following example of two spins \cite{AudenaertPlenio06}: Suppose one measures
the correlation
$\langle\hat{\sigma}_1^z\hat{\sigma}_2^z\rangle-\langle\hat{\sigma}_1^z\rangle\langle\hat{\sigma}_2^z\rangle$ and obtains the result $-1$. This measurement is consistent with
{\em both} the maximally entangled state $|\psi\rangle=(|\!\!\uparrow\downarrow\rangle-|\!\!\downarrow\uparrow\rangle)/\sqrt{2}$ {\em and} the separable state $\hat{\varrho}=(|\!\!\uparrow\downarrow\rangle\langle \uparrow\downarrow\!\!|+|\!\!\downarrow\uparrow\rangle\langle \downarrow\uparrow\!\!|)/2$. Hence, without further assumptions, one may not decide whether the state in the laboratory is entangled or not. A possible
assumption may be that the system is in thermal equilibrium at some known temperature {\em and} that the Hamiltonian that governs the system is known
precisely. But obtaining knowledge about the Hamiltonian experimentally is
even harder than to obtain the state itself! A technique to decide without
a doubt wether  entanglement is contained in a given system should hence not
rely on knowing the Hamiltonian, it should, in fact, not
rely on any kind of knowledge about the system other than measured data, but
be able to quantify entanglement by just relying on measured observables.

Here we present a scheme to quantify entanglement in condensed matter systems
that fulfils all the above requirements and relies only on measurements that
are already available: Neutron scattering from spin systems and time-of-flight
imaging of cold atoms. Hence, we show that it is possible to directly---without
any assumptions---measure entanglement in many-body systems. To this end we
exploit the substantial body of work concerning the characterization (which
states are entangled), quantification (how much of it do we have) and
verification (on the basis of simple measurements we need to answer the
previous questions) of entanglement (see \cite{PlenioV07} for a tutorial
review and \cite{Horoedecki09} for a advanced and very comprehensive review)
that has been established in quantum information science. More precisely,
we combine methods for determining the presence of entanglement in quantum
many-body system \cite{Bruss09,wunderlich} with proposals for the quantitative
verification of entanglement in few body quantum systems \cite{AudenaertPlenio06,Science,Eisert07,Guehne07} to achieve an experimentally
accessible method for measuring entanglement.

\section{Spin systems: Neutron scattering}

One of the standard tools to analyze condensed matter samples is neutron scattering, see, e.g., Ref.~\cite{special_issue}. The deflected neutrons
carry information about both the structural and magnetic properties of
the sample, which can be read of the differential scattering cross section
\cite{cross_section}. The Fourier transform of the magnetic cross section
gives access to spin correlations in reciprocal space such as the positive
hermitian observable
$\hat{S}(\vec{q})=\sum_{\alpha=x,y,z}\hat{S}_\alpha(\vec{q})$, where
\begin{equation}
\label{spin_observe}
    \hat{S}_\alpha(\vec{q})=\frac{1}{M}\sum_{\vec{i},\vec{j}}\me^{\mi \vec{q}\cdot(\vec{r}_{\vec{i}}-\vec{r}_{\vec{j}})}\hat{\sigma}_{\vec{i}}^{\alpha}\hat{\sigma}_{\vec{j}}^{\alpha},
\end{equation}
$\hat{\sigma}_{\vec{i}}^{x}=(\begin{smallmatrix} 0&1\\ 1&0 \end{smallmatrix})$, $\hat{\sigma}_{\vec{i}}^{y}=(\begin{smallmatrix} 0&-\mi\\ \mi&0 \end{smallmatrix})$, $\hat{\sigma}_{\vec{i}}^{z}=(\begin{smallmatrix} 1&0\\ 0&-1 \end{smallmatrix})$  are Pauli spin matrices acting on lattice site $\vec{i}$  located at $\vec{r}_{\vec{i}}$, and $M$ is the total number of spins, so the number of lattice sites.
Analogous observables can be obtained for spin systems realized in ion traps or cold atoms in optical lattices, in which one can directly measure spin correlations by light scattering \cite{deVega}.

In the following we will show how a measurement of observables of the type in Eq.~(\ref{spin_observe})   alone is sufficient to quantify the entanglement contained in the sample.
We will consider systems comprised of spin $1/2$ particles (generalizations to
higher spins are entirely straightforward) on a lattice and set out to derive
a lower bound to the entanglement that is consistent with the measurement data.
It will turn out that it is possible to find a lower bound that is a simple
function of the static structure factor $\langle \hat{S}(\vec{q})\rangle$. Hence, for this, no assumptions about
the state are necessary, in particular, no knowledge about the Hamiltonian is required.

In the following we present definitions and a derivation that will lead
to the central result in Eq.~(\ref{spin_central_result}), which provides
a lower bound on the entanglement of any spin-state under investigation
that can be used directly on experimental data. Several entanglement measures
may be expressed in the form \cite{fernando,PlenioV07}
\begin{equation}
\label{1}
\mathcal{E}_{\mathcal{C}}(\hat{\varrho})=\max\bigl\{0,-\min_{\hat{W}\in\mathcal{W}\cap\mathcal{C}}\text{tr}[\hat{W}\hat{\varrho}]\bigr\},
\end{equation}
where $\mathcal{W}$ denotes the set of hermitian operators that fulfil $\langle\hat{W}\rangle\ge 0$ for separable states (i.e., the set of {\em entanglement witnesses} \cite{witness_review}) and $\mathcal{C}$ distinguishes the quantities: E.g.,
if $\mathcal{C}$ is the set of operators $\hat{W}$ fulfilling $\langle\hat{W}\rangle\le 1$ for separable states
then $\mathcal{E}_{\mathcal{C}}$ is the robustness of entanglement \cite{robustness}, for $\mathcal{C}=\{\hat{W}\in\mathcal{W}\,|\,\id-\hat{W}\ge 0\}$, $\mathcal{E}_{\mathcal{C}}$ measures the generalized robustness of entanglement \cite{gen_robustness}, and
if $\mathcal{C}=\{\hat{W}\in\mathcal{W}\,|\,\id+\hat{W}\ge 0\}$ then $\mathcal{E}_{\mathcal{C}}$ is equal to the best separable approximation  \cite{bsa}. In fact, $E_{\mathcal{C}_{n,m}}$, $\mathcal{C}_{n,m}:=\{\hat{W}\in\mathcal{W}\,|\,-n\id\le\hat{W}\le m\id\}$, is an entanglement monotone
for every $n,m\ge 0$ \cite{fernando}. One can now exploit the fact that for any choice $\hat{W}\in\mathcal{W}\cap\mathcal{C}$ one obtains a lower bound to $\mathcal{E}_{\mathcal{C}}$,
\begin{equation}
\mathcal{E}_{\mathcal{C}}(\hat{\varrho})\ge \max\bigl\{0,-\text{tr}[\hat{W}\hat{\varrho}]\bigr\}\text{ for all } \hat{W}\in\mathcal{W}\cap\mathcal{C}.
\end{equation}
Given this expression, it is possible to arrive at lower bounds to $\mathcal{E}_{\mathcal{C}}$
by simply constructing operators $\hat{W}\in\mathcal{W}\cap\mathcal{C}$ that are functions of observables that are within experimental reach. This works of course for {\em any} spin system and {\em any} observable.

In the following, we will focus on the best separable approximation and the observable $\hat{S}$. Consider the operator
\begin{equation}
\hat{W}(\vec{q})=
\frac{1}{2}\hat{S}(\vec{q})-\id,
\end{equation}
for which we now show that $\hat{W}\in\mathcal{W}\cap\mathcal{C}$. We find $\hat{W}(\vec{q})+\id\ge 0$ and for a product state $\hat{\varrho}=\otimes_{\vec{i}}\hat{\varrho}_{\vec{i}}$, we have
\begin{equation}
\begin{split}
\langle\hat{W}(\vec{q})\rangle&=\frac{1}{2M}\sum_{\vec{i},\alpha}\left(1-\langle\hat{\sigma_{\vec{i}}}^\alpha\rangle^2\right)-1\\
&\hspace{1.5cm}+\frac{1}{2M}\sum_{\alpha}\bigl|\sum_{\vec{i}}\me^{\mi \vec{q}\cdot\vec{r}_{\vec{i}}}\langle\hat{\sigma}_{\vec{i}}^{\alpha}\rangle\bigr|^2,
\end{split}
\end{equation}
which is non-negative as the last term is and the first term may be bounded by using the uncertainty relation $\sum_{\alpha}(1-\langle\hat{\sigma}_{\vec{i}}^\alpha\rangle^2)\ge 2$. Hence, $\hat{W}\in\mathcal{W}\cap\mathcal{C}$, i.e.,
for every state $\hat{\varrho}$ and every $\vec{q}$, the quantity
\begin{equation}
\label{spin_central_result}
E(\vec{q})=\max\big\{0,1-
\frac{1}{2M}\sum_{\vec{i},\vec{j},\alpha}\me^{\mi \vec{q}\cdot(\vec{r}_{\vec{i}}-\vec{r}_{\vec{j}})}\langle\hat{\sigma}_{\vec{i}}^{\alpha}\hat{\sigma}_{\vec{j}}^{\alpha}\rangle\big\}
\end{equation}
provides a lower bound to the $M$-partite entanglement (as measured in terms of the best separable approximation) contained in $\hat{\varrho}$.
Similar bounds may be derived for all entanglement measures that fall into the general framework of  Eq.\ (\ref{1}).

At this point, we would like to emphasise again that $E(\vec{q})$ gives a lower bound to the entanglement for {\em any} state on the lattice -- irrespective of
how it has been prepared, what the temperature is, or what the underlying Hamiltonian of the system might be.
\begin{figure}
\includegraphics[width=0.9\columnwidth]{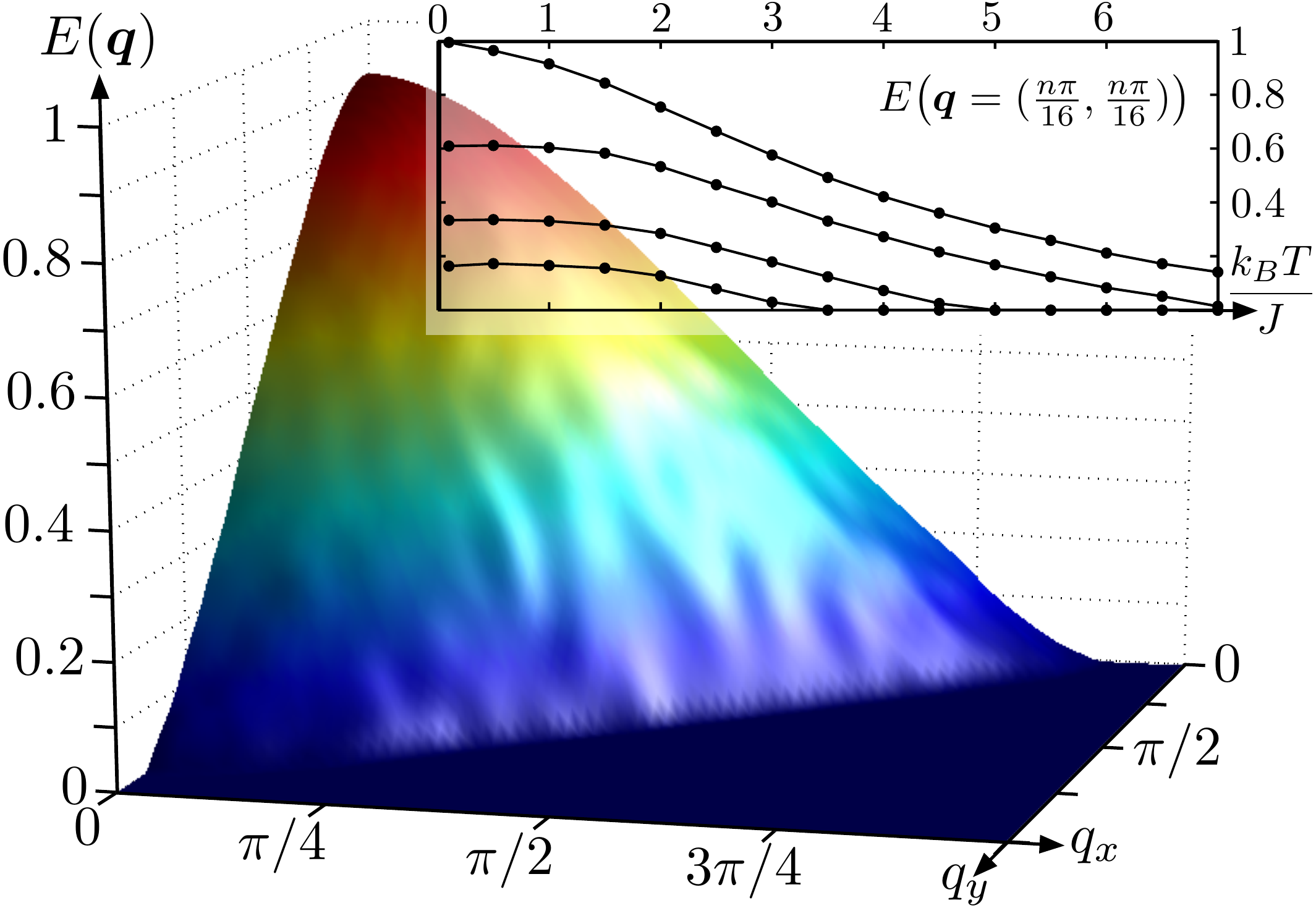}
\caption{\label{fig_spin}
Lower bound $E$ on the entanglement---as measured in terms of the best separable approximation (BSA)---for a thermal state $\hat{\varrho}=\exp(-\beta\hat{H})/Z$, $\beta J= J/k_BT=1$, of the Heisenberg model in Eq.~(\ref{heisenberg}). For every $\vec{q}$, $E(\vec{q})$ provides a lower bound to the BSA. The square lattice with open boundary conditions and lattice constant $a=1$ has $30\times 30$ lattice sites and $\langle\hat{S}(\vec{q})\rangle$ was obtained using the generalized directed loop quantum Monte Carlo algorithm \cite{loop} of the ALPS package \cite{ALPS}. Inset shows $E(\frac{n\pi}{16},\frac{n\pi}{16})$, $n=0, 4, 6, 7$ (top to bottom), as a function of the temperature. Lines are a guide to the eye. Note that the BSA is upper bounded by unity, a bound that $E(\vec{q})$ saturates at low temperatures and $\vec{q}=\vec{0}$.
}
\end{figure}

As an example, we consider thermal states of the antiferromagnetic Heisenberg model
\begin{equation}
\label{heisenberg}
\hat{H}=J\sum_{\alpha=x,y,z}\sum_{\langle i,j\rangle}\hat{\sigma}_i^\alpha\hat{\sigma}_i^\alpha
\end{equation}
on a square lattice. This model has been analyzed in great detail in the literature (see Ref.~\cite{heisenberg_review} for a review) using several analytical and numerical techniques. In Ref.~\cite{aeppli}
the two-dimensional Heisenberg antiferromagnet copper deuteroformate tetradeurate has been analyzed experimentally using extensive neutron scattering measurements and it has been suggested---under the assumption that the system is indeed described by the Heisenberg Hamiltonian with known coupling constants  and by a comparison of a classical to a quantum description---that entanglement is present in this system.  Using $E(\vec{q})$, the presence of entanglement can not only be
confirmed but, as it is a lower bound to the best separable approximation, also
quantified. In Fig.~\ref{fig_spin}, we show $E(\vec{q})$ for a thermal state
of the Heisenberg model at different values of $J\beta$ as obtained from a quantum Monte Carlo
simulation using the generalized loop algorithm \cite{loop} of the ALPS library \cite{ALPS}. The plot shows that entanglement
is present up to fairly high temperatures, i.e., measuring entanglement is well within experimental reach
(in \cite{aeppli}, e.g., the sample was at a temperature of $1.5\,$K and the
data well fitted by $J=6.19\,$meV, i.e., $k_BT/J=0.02$). In addition, the plot exemplifies the quality of our bound: The best separable approximation is upper bounded by unity and at low temperatures $E(\vec{0})\approx 1$. $E(\vec{q})$ also scales properly with the system size: For the ground state of the Heisenberg model it is known that $S(\vec{q})\sim |\vec{q}|$ for small $|\vec{q}|$ \cite{heisenberg_review}, i.e., $E(\vec{q})-1\sim |\vec{q}|$, see also footnote \cite{system_size}.

\section{Bosons in optical lattices: Time-of-flight densities}
A standard measurement in the context of ultracold atoms is the following: One switches off all potentials, allows the atom cloud to expand freely and then takes an absorption image of the atoms, which reveals the {\em velocity}, {\em quasi-momentum}, or {\em time-of-flight distribution} of the atoms before the expansion. This technique was used to show Bose-Einstein
condensation into the lowest-momentum state \cite{BEC}, to demonstrate the Mott insulator -- superfluid transition of bosons in optical lattices \cite{Mott}, and to observe Fermi surfaces of fermions in optical lattices \cite{FermiSurface} to name just a few. We focus on the situation in which bosons of mass $m$ are kept in an optical lattice with lattice constant $a$. After a time of flight $t$, the density of the atoms reads (see, e.g., Refs.~\cite{stringari_qmd, bloch_qmd})
\begin{equation}
n(\vec{r})=\sum_{\vec{i},\vec{j}}f_{\vec{i},\vec{j}}(\vec{k}=\tfrac{m\vec{r}}{\hbar t})\langle\hat{b}_{\vec{i}}^\dagger\hat{b}_{\vec{j}}\rangle,
\end{equation}
where
\begin{equation}
\begin{split}
f_{\vec{i},\vec{j}}(\vec{k})&=(\tfrac{m}{\hbar t})^3|w(\vec{k})|^2\me^{\mi (\vec{k}(\vec{r}_{\vec{i}}-\vec{r}_{\vec{j}})+\frac{m}{2\hbar t}(\vec{r}_{\vec{j}}^2-\vec{r}_{\vec{i}}^2))},
\end{split}
\end{equation}
$w(\vec{k})$ is the Fourier-transform of the Wannier function centred at zero, and $\hat{b}_{\vec{i}}$ annihilates a boson at site $\vec{i}$ located at $\vec{r}_{\vec{i}}$.
The resulting absorption image is then the integral along the optical axis, say, the $z$-direction, of this density, i.e.,
\begin{equation}
n(x,y)=\sum_{\vec{i},\vec{j}}f_{\vec{i},\vec{j}}(x,y)\langle\hat{b}_{\vec{i}}^\dagger\hat{b}_{\vec{j}}\rangle=:\langle\hat{n}(x,y)\rangle,
\end{equation}
where $f_{\vec{i},\vec{j}}(x,y)=\int\!\md z\,f_{\vec{i},\vec{j}}(\tfrac{\hbar t\vec{r}}{m})$, $f_{\vec{i},\vec{i}}(x,y)=:f(x,y)$ . We now set out to show that
\begin{equation}
\label{bose_bound}
E(x,y)=\max\big\{0,\langle\hat{N}\rangle-\frac{n(x,y)}{f(x,y)}\big\}
\end{equation}
provides a lower bound to the entanglement contained in the state in the laboratory, which constitutes the main result of this section. Here, $\langle\hat{N}\rangle=\sum_{\vec{i}}\langle\hat{b}_{\vec{i}}^\dagger\hat{b}_{\vec{i}}\rangle$ is the expected total number of atoms.

As we are concerned with massive particles, we will, in the following restrict the state space to states $\hat{\varrho}$ that have
a finite mean number of particles, $\text{tr}[\hat{\varrho}\hat{N}]<\infty$, and commute with the particle number operator $\hat{N}$. In other words, we are concerned with states respecting the particle-number superselection rule (SSR) -- the only physical states allowed in this setting of indistinguishable massive particles \cite{ssr}. These states are of the form $\hat{\varrho}=\sum_{N=0}^\infty\hat{P}_N\hat{\varrho}\hat{P}_N=:\bigoplus_{N=0}^\infty\hat{\varrho}_N$, where $\hat{P}_N$ projects on the sector with constant particle number $N$. The SSR also restricts the allowed physical operations to operations commuting with $\hat{N}$ \cite{ssr_operations}.
Consider now
\begin{equation}
\label{bose_monotone}
\mathcal{E}(\hat{\varrho})=\max\bigl\{0,-\sum_{N=0}^\infty\min_{\hat{W}\in\mathcal{C}_N}\text{tr}[\hat{\varrho}_N\hat{W}]\bigr\},
\end{equation}
where $\mathcal{C}_N$ is the set of hermitian operators $\hat{W}$ acting on the subspace of constant particle number $N$
that fulfil $cN\pm\hat{W}\ge 0$ for some constant $c>0$ independent of $N$ and $\text{tr}[\hat{\varrho}_N\hat{W}]\ge 0$ for separable $\hat{\varrho}_N$. In the Appendix we show that $\mathcal{E}$
is an entanglement monotone under LOCC operations that preserve the {\em total} number of particles, i.e., that commute with $\hat{N}$ (and hence it is also an entanglement monotone under SSR-LOCC operations -- LOCC operations that preserve the {\em local} particle number; for a discussion of entanglement under SSR see Ref.~\cite{ssr}).

We now show that $\hat{W}_N=\hat{P}_N(\hat{n}/f(x,y)-\hat{N})\hat{P}_N\in\mathcal{C}_N$. To this end let $|\psi\rangle$ be a state vector on the subspace of constant particle number $N$. Then, with $M$ being the number of lattices sites, we find
\begin{equation}
\begin{split}
\frac{\langle\psi|
\hat{n}
|\psi\rangle}{f(x,y)}&\le
\sum_{\vec{i},\vec{j}}|\langle\psi|\hat{b}_{\vec{i}}^\dagger\hat{b}_{\vec{j}}|\psi\rangle|\\
&\le \sum_{\vec{i},\vec{j}}\sqrt{\langle\psi|\hat{b}_{\vec{i}}^\dagger\hat{b}_{\vec{i}}|\psi\rangle}
\sqrt{\langle\psi|\hat{b}_{\vec{j}}^\dagger\hat{b}_{\vec{j}}|\psi\rangle}\\
&\le
\sum_{\vec{i},\vec{j}}\tfrac{\langle\psi|\hat{b}_{\vec{i}}^\dagger\hat{b}_{\vec{i}}|\psi\rangle+\langle\psi|\hat{b}_{\vec{j}}^\dagger\hat{b}_{\vec{j}}|\psi\rangle}{2}
= NM,
\end{split}
\end{equation}
i.e., $MN\pm\hat{W}_N\ge 0$. Furthermore, for separable $\hat{\varrho}_N$, we find
for $\vec{i}\ne\vec{j}$ that $\text{tr}[\hat{\varrho}_N\hat{b}_{\vec{i}}^\dagger\hat{b}_{\vec{j}}]=0$
and hence
\begin{equation}
\begin{split}
\text{tr}[\hat{\varrho}_N\hat{n}]
=f(x,y)\sum_{\vec{i}}
\text{tr}[\hat{\varrho}_N\hat{b}_{\vec{i}}^\dagger\hat{b}_{\vec{i}}]
=f(x,y)N,
\end{split}
\end{equation}
i.e., $\text{tr}[\hat{\varrho}_N\hat{W}_N]\ge 0$.
Hence, $\hat{W}_N\in\mathcal{C}_N$,  which implies that
\begin{equation}
-\min_{\hat{W}\in\mathcal{C}_N}\text{tr}[\hat{\varrho}_N\hat{W}]\ge
\text{tr}[\hat{\varrho}_N(N-\hat{n}/f(x,y))]
\end{equation}
 and thus, for all $x,y$, the quantity $E(x,y)$ in Eq.~(\ref{bose_bound})
provides a lower bound to the $M$-partite entanglement (as measured in terms of $\mathcal{E}$) available in the system. Here, $n(x,y)=\langle\hat{n}(x,y)\rangle$ is obtained in standard time-of-flights measurements and $E$ is a lower bound for {\em any} state on the lattice.
\begin{figure}
\includegraphics[width=\columnwidth]{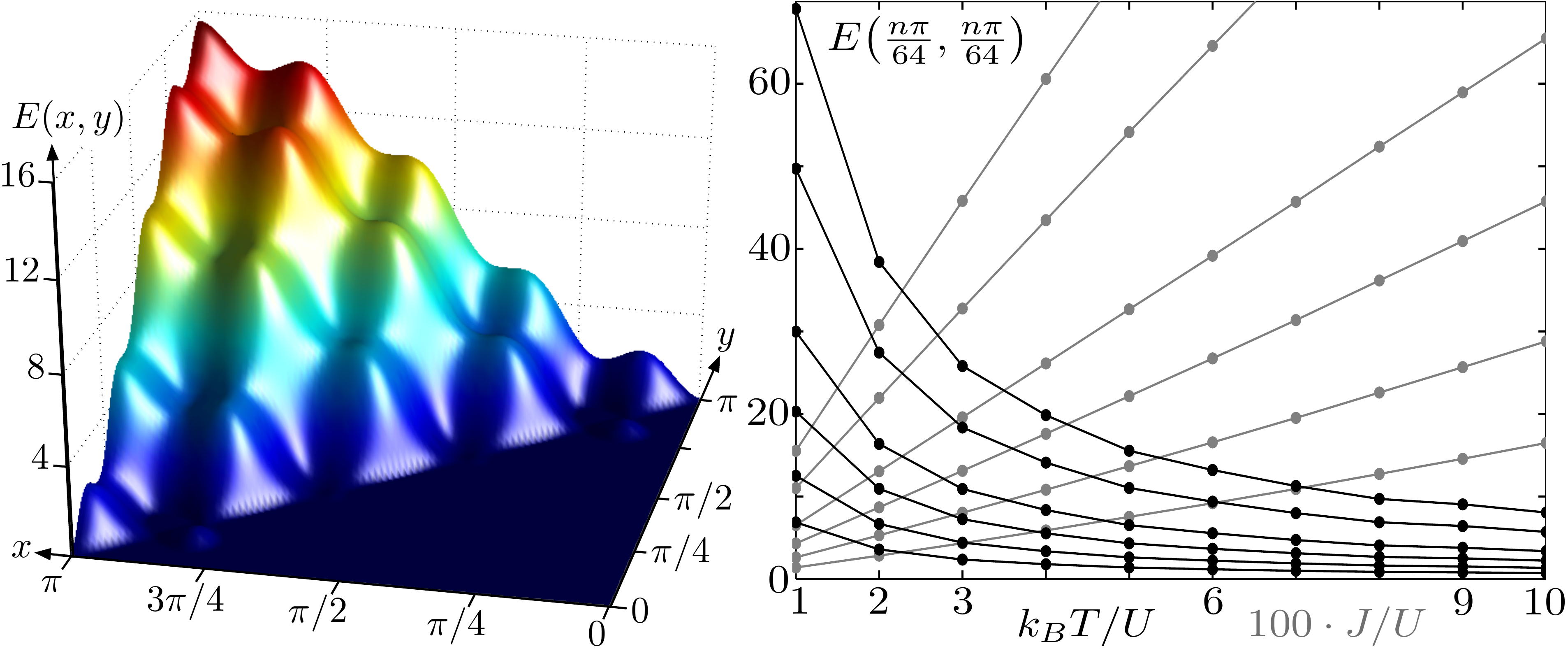}
\caption{\label{fig_bose}
Lower bound $E(x,y)$ on the entanglement for a thermal state $\hat{\varrho}=\exp(-\hat{H}/k_BT)/Z$ with constant filling factor $\langle\hat{n}_{\vec{i}}\rangle=1$ of the Bose-Hubbard model in Eq.~(\ref{bose_hubbard}). The three-dimensional cubic lattice with periodic boundary conditions
and lattice constant $a=1$ has $10\times 10\times 10$ lattice sites and $\langle\hat{n}(x,y)\rangle$ was obtained using the same numerical code as for Fig.~\ref{fig_spin}. Left plot shows $E(x,y)$ as in Eq.~(\ref{bose_bound}) for
$\beta U=1/5$, $J/U=0.01$. Right plot shows $E(\frac{n\pi}{64},\frac{n\pi}{64})$, $n=64, 48, 44, 36, 34, 33$ (top to bottom) as a function of the temperature (black) and of the tunnelling amplitude $J/U$ (gray). Lines are guides to the eye.
}
\end{figure}

As an example, we consider thermal states of the Bose-Hubbard model on a three dimensional cubic lattice,
\begin{equation}
\label{bose_hubbard}
\hat{H}=-J\sum_{\langle\vec{i},\vec{j}\rangle}\hat{b}_{\vec{i}}^\dagger\hat{b}_{\vec{j}}+\frac{U}{2}\sum_{\vec{i}}\hat{n}_{\vec{i}}(\hat{n}_{\vec{i}}-1)-\mu
\sum_{\vec{i}}\hat{n}_{\vec{i}},
\end{equation}
where $\hat{n}_i=\hat{b}^\dagger_{\vec{i}}\hat{b}_{\vec{i}}$, $\langle\cdot,\cdot\rangle$ denotes summation over nearest neighbours, $J$ accounts for tunnelling of atoms between adjacent sites, $U$ is the strength of the on-site repulsion of atoms, and the chemical potential $\mu$ controls the particle number. Bosons in deep optical lattices are (up to an harmonic confinement) well described by this model, which displays a quantum phase transition from a Mott insulator (small $J/U$) to superfluid (large $J/U$) that was observed in \cite{Mott} via the interference pattern displayed in $\langle\hat{n}(x,y)\rangle$.
In Fig.~\ref{fig_bose}, we show $E(x,y)$ for a thermal state of the Bose-Hubbard model as obtained from a quantum Monte Carlo computation using the ALPS library~\cite{ALPS}. We can see that $E(x,y)$ increases linearly with $J/U$ and stays finite up to high temperatures. Furthermore, in experiments the atoms are harmonically trapped leading to an extension of the atoms over the lattice that is on the order of the system size considered here.
Hence,
quantifying entanglement in these systems is already well within experimental reach.

\section{Conclusion}
We have derived lower bounds to the entanglement contained in lattice systems.
These lower bounds are a simple function of routinely measured observables and do not require any additional information about the system. This makes not only the verification but also the quantification
of entanglement in condensed matter samples possible. In other words, without making any assumptions (such as the temperature, the Hamiltonian that governs the system, the way the state was created),
entanglement can be directly measured using only measurements that already belong to the toolbox
for the analysis of quantum many-body systems. The presented schemes straightforwardly generalize to other many-body systems and observables and we foster the hope that they will see  direct application to experimentally realized situations and inspire further generalizations.

\section{Acknowledgements}
We would like to acknowledge fruitful discussions with F.~Brand{\~a}o and thank S.~Wessel and M.~Troyer for helpful comments on the ALPS library.
This work is supported by the EU Integrated project Q-ESSENCE, the EU STREP projects CORNER and HIP and the Alexander-von-Humboldt Foundation. Computations were performed on the bwGRiD \cite{grid}. The first and third author contributed equally to this work.

\section{Appendix}
We set out to show that on the set of states $\hat{\varrho}$ that respect the particle number superselection rule (i.e., are of the form $
\hat{\varrho}=\sum_{N=0}^\infty\hat{P}_N\hat{\varrho}\hat{P}_N$) and have a finite mean number of particles, $\text{tr}[\hat{\varrho}\hat{N}]<\infty$, the quantity $\mathcal{E}$ as in Eq.~(\ref{bose_monotone}) is an entanglement monotone under LOCC operations that preserve the total particle number. First, we note that $\mathcal{E}$ is well defined:
The set $\mathcal{C}_N$ is a subset of hermitian operators on $\mathcal{N}_N$, which we denote by $\mathcal{H}_N$,
and the trace is understood to be over $\mathcal{N}_N$. Then, for given $N$ and $\hat{\varrho}_N$ the mapping $\mathcal{H}_N\rightarrow \text{tr}[\hat{\varrho}_N\,\cdot\,]$ is continuous, i.e., the minimum is attained on $\mathcal{C}_N$  as $\mathcal{C}_N$ is compact:
$\mathcal{C}_N=\mathcal{A}_N\cap\mathcal{S}_N$, where
\begin{equation}
\mathcal{A}_N=\bigl\{\hat{W}\in\mathcal{H}_N\,\big|\, -cN\le\hat{W}\le cN\bigr\},
\end{equation}
is compact and the complement of
\begin{equation}
\mathcal{S}_N=\bigl\{\hat{W}\in\mathcal{H}_N\,\big|\, \text{tr}[\hat{\varrho}_N\hat{W}]\ge 0\text{ for separable }\hat{\varrho}_N\bigr\}
\end{equation}
is open in $\mathcal{H}_N$: Let $\hat{W}\in \mathcal{H}_N$ but not in $\mathcal{S}_N$. Then there exists a non-separable $\hat{\varrho}_N$ such that
$\text{tr}[\hat{\varrho}_N\hat{W}]/2= -\epsilon <0$. Then for all $\hat{W}^\prime\in \mathcal{H}_N$ with $\|\hat{W}^\prime-\hat{W}\|<\epsilon$, we have
$\text{tr}[\hat{\varrho}_N\hat{W}^\prime]=\text{tr}[\hat{\varrho}_N(\hat{W}^\prime-\hat{W})]+\text{tr}[\hat{\varrho}_N\hat{W}]\le \|\hat{W}^\prime-\hat{W}\|-2\epsilon < -\epsilon <0$. Hence, the minimum is attained and, for given $\hat{\varrho}_N$ denoting the minimizer by $\hat{W}_{\hat{\varrho}_N}$, we may write
\begin{equation}
E(\hat{\varrho})=\max\Bigl\{0,-\sum_{N=0}^\infty\text{tr}[\hat{\varrho}_N\hat{W}_{\hat{\varrho}_N}]\Bigr\}.
\end{equation}
Finally, the series is absolutely convergent:
\begin{equation}
-c N\text{tr}[\hat{\varrho}_N]\le \text{tr}[\hat{\varrho}_N\hat{W}]\le cN\text{tr}[\hat{\varrho}_N]
\end{equation}
for all $\hat{W}\in \mathcal{C}_N$, i.e., $|\text{tr}[\hat{\varrho}_N\hat{W}]|\le cN\text{tr}[\hat{\varrho}_N]$ for all $\hat{W}\in \mathcal{C}_N$, and therefore
\begin{equation}
\sum_{N=0}^\infty|\text{tr}[\hat{\varrho}_N\hat{W}_{\hat{\varrho}_N}]|\le c\sum_{N=0}^\infty N\text{tr}[\hat{\varrho}_N]
=c\,\text{tr}[\hat{\varrho}\hat{N}]<\infty.
\end{equation}
Hence, $\mathcal{E}$ is well defined.

Now let $\hat{A}_k$, $k=1,2,\dots$, be Kraus operators of the form $\hat{A}_k=\sum_{N=0}^\infty \hat{A}_N^k$ (i.e., we only allow operations that preserve the total number of particles) with
$\sum_{k}\hat{A}_k^\dagger\hat{A}_k\le \id$ and $\hat{A}_k=\bigotimes_{s}\hat{A}_s^k$, where the direct product refers to some partition of the system. For a given state $\hat{\varrho}=\bigoplus_{N=0}^\infty\hat{\varrho}_N$, denote
\begin{equation}
\begin{split}
\hat{R}_k&=\hat{A}_k\hat{\varrho}\hat{A}_k^\dagger=\bigoplus_{N=0}^\infty\hat{A}_N^k\hat{\varrho}_N(\hat{A}_N^k)^\dagger,
\end{split}
\end{equation}
$p_k=\text{tr}[\hat{R}_k]$, $K=\{k\,|\,p_k>0\}$, and for $k\in K$ write
\begin{equation}
\hat{\varrho}_k=\hat{R}_k/p_k=\frac{1}{p_k}\bigoplus_{N=0}^\infty \hat{A}_N^k\hat{\varrho}_N(\hat{A}_N^k)^\dagger=:\bigoplus_{N=0}^\infty\hat{\varrho}_N^k.
\end{equation}
Then, with $K^\prime=\{k\in K\,|\,\sum_{N=0}^\infty\text{tr}[\hat{\varrho}^k_N\hat{W}_{\hat{\varrho}^k_N}]<0\}$,
\begin{equation}
\begin{split}
\sum_{k\in K}p_kE(\hat{\varrho}_k)&=\sum_{k\in K}p_k\max\left\{0,-\sum_{N=0}^\infty\text{tr}[\hat{\varrho}^k_N\hat{W}_{\hat{\varrho}^k_N}]\right\}\\
&=-\sum_{k\in K^\prime}p_k\sum_{N=0}^\infty\text{tr}[\hat{\varrho}^k_N\hat{W}_{\hat{\varrho}^k_N}]\\
&=-\sum_{k\in K^\prime}\sum_{N=0}^\infty\text{tr}[ \hat{\varrho}_N(\hat{A}_N^k)^\dagger  \hat{W}_{\hat{\varrho}^k_N}\hat{A}_N^k].
\end{split}
\end{equation}
We already know that $\sum_{N=0}^\infty\text{tr}[\hat{\varrho}^k_N\hat{W}_{\hat{\varrho}^k_N}]$ converges absolutely. We also have
\begin{equation}
\begin{split}
\sum_{k\in K^\prime}p_k|\text{tr}[\hat{\varrho}^k_N\hat{W}_{\hat{\varrho}^k_N}]|&=\sum_{k\in K^\prime}|\text{tr}[ \hat{A}_N^k\hat{\varrho}_N(\hat{A}_N^k)^\dagger  \hat{W}_{\hat{\varrho}^k_N}]|\\
&\le cN\sum_{k}\text{tr}[ \hat{A}_N^k\hat{\varrho}_N(\hat{A}_N^k)^\dagger]\\
&= cN\text{tr}[ \hat{\varrho}_N\bigl(\sum_{k}(\hat{A}_N^k)^\dagger\hat{A}_N^k\bigr)],
\end{split}
\end{equation}
which is upper bounded by $cN\text{tr}[ \hat{\varrho}_N]$,
i.e.,
\begin{equation}
\begin{split}
\sum_{N=0}^\infty\sum_{k\in K^\prime}p_k|\text{tr}[\hat{\varrho}^k_N\hat{W}_{\hat{\varrho}^k_N}]|&\le
 c\sum_{N=0}^\infty N\text{tr}[ \hat{\varrho}_N]=c\text{tr}[\hat{\varrho}\hat{N}],
\end{split}
\end{equation}
which is finite. Hence, we may interchange the sums to find
\begin{equation}
\begin{split}
\sum_{k\in K}p_kE(\hat{\varrho}_k)&=-\sum_{N=0}^\infty\text{tr}\bigl[ \hat{\varrho}_N\bigl(\sum_{k\in K^\prime}(\hat{A}_N^k)^\dagger  \hat{W}_{\hat{\varrho}^k_N}\hat{A}_N^k\bigr)\bigr]\\
&\underset{(*)}{\le}
-\sum_{N=0}^\infty\text{tr}[ \hat{\varrho}_N\hat{W}_{\hat{\varrho}_N}]
\le E(\hat{\varrho}),
\end{split}
\end{equation}
where $(*)$ holds if $\sum_{k\in K^\prime}(\hat{A}_N^k)^\dagger  \hat{W}_{\hat{\varrho}^k_N}\hat{A}_N^k$ is an element of $\mathcal{C}_N$,
which we now set out to show. As $\sum_k\hat{A}_k^\dagger\hat{A}_k\le\id$, we also have $\sum_k(\hat{A}_N^k)^\dagger\hat{A}_N^k\le \id$,
i.e., bounding positive semi-definite operators by zero,
\begin{equation}
\begin{split}
&\hspace{-1cm}cN\id\pm\sum_{k\in K^\prime}(\hat{A}_N^k)^\dagger  \hat{W}_{\hat{\varrho}^k_N}\hat{A}_N^k\\
&\ge\sum_k(\hat{A}_N^k)^\dagger cN\hat{A}_N^k
\pm\sum_{k\in K^\prime}(\hat{A}_N^k)^\dagger  \hat{W}_{\hat{\varrho}^k_N}\hat{A}_N^k\\
&\ge \sum_{k\in K^\prime}(\hat{A}_N^k)^\dagger (cN\id\pm \hat{W}_{\hat{\varrho}^k_N})\hat{A}_N^k\ge 0,
\end{split}
\end{equation}
i.e., $\sum_{k\in K^\prime}(\hat{A}_N^k)^\dagger  \hat{W}_{\hat{\varrho}^k_N}\hat{A}_N^k\in\mathcal{A}_N$.
Now let $\hat{\varrho}_N$ be separable. Then
\begin{equation}
\begin{split}
\text{tr}\bigl[\hat{\varrho}_N\bigl(\sum_{k\in K^\prime}(\hat{A}_N^k)^\dagger  \hat{W}_{\hat{\varrho}^k_N}\hat{A}_N^k\bigr)\bigr]\hspace{2cm}\\
=\sum_{k\in K^\prime}\text{tr}\left[\hat{A}_N^k\hat{\varrho}_N(\hat{A}_N^k)^\dagger  \hat{W}_{\hat{\varrho}^k_N}\right],
\end{split}
\end{equation}
where each summand is non-negative: The $\hat{A}_N^k\hat{\varrho}_N(\hat{A}_N^k)^\dagger$ are
separable as the  $\hat{A}_N^k$ are local operations and $\hat{\varrho}_N$ is separable. Therefore, as $\hat{W}_{\hat{\varrho}^k_N}\in\mathcal{S}_N$ we have
\begin{equation}
\text{tr}\left[\hat{A}_N^k\hat{\varrho}_N(\hat{A}_N^k)^\dagger  \hat{W}_{\hat{\varrho}^k_N}\right]\ge 0,
\end{equation}
i.e., finally, $\sum_{k\in K^\prime}(\hat{A}_N^k)^\dagger  \hat{W}_{\hat{\varrho}^k_N}\hat{A}_N^k\in\mathcal{S}_N$.


\begin{thebibliography}{99}
\bibitem{system_size} We have checked the scaling of $E(\vec{q})$ with the system size. E.g., at $J\beta=1$, we found $E_{30}(\vec{0})-E_{10}(\vec{0})=0.0055$ and $E_{30}(\vec{0})-E_{20}(\vec{0})=0.0021$, where the index indicates the linear size of the square lattice.
\end{thebibliography}
\end{document}